\documentclass[journal=jpclcd,manuscript=letter]{achemso}

\usepackage[version=3]{mhchem}
\usepackage{caption}
\usepackage{float}
\usepackage{geometry}
\usepackage{setspace}
\usepackage{array}
\usepackage{multirow}
\usepackage{adjustbox}
\usepackage{longtable}
\usepackage{graphicx}
\usepackage{epstopdf}
\usepackage{subcaption}
\usepackage[usenames,dvipsnames]{xcolor}
\usepackage{comment}

\usepackage{makecell}

\newcolumntype{?}{!{\vrule width 1.2pt}}

\author{Shashikant Kumar}
\affiliation[Georgia Tech]{College of Engineering, Georgia Institute of Technology, Atlanta, GA 30332, USA}
\author{David Codony}
\affiliation{Laboratori de C\`{a}lcul Num\`{e}ric, Universitat Polit\`{e}cnica de Catalunya, Barcelona, E-08034, Spain}
\author{Irene Arias}
\affiliation{Laboratori de C\`{a}lcul Num\`{e}ric, Universitat Polit\`{e}cnica de Catalunya, Barcelona, E-08034, Spain}
\altaffiliation{Centre Internacional de Metodes Numèrics en Enginyeria (CIMNE), 08034 Barcelona, Spain}
\author{Phanish Suryanarayana}
\email{phanish.suryanarayana@ce.gatech.edu}
\affiliation[Georgia Tech]{College of Engineering, Georgia Institute of Technology, Atlanta, GA 30332, USA}

\title{Transversal flexoelectric coefficients for  fifty select atomic monolayers from first principles}
\keywords{Flexoelectric effect, Density Functional Theory, Two-dimensional materials, Electromechanical coupling, Dipole moment, Strain gradient}

\begin{document}

%\makeatletter
%\setlength\acs@tocentry@height{5 cm}
%\setlength\acs@tocentry@width{3.75 cm}
%\makeatother

\setlength{\fboxrule}{0 pt}
\begin{tocentry}
\includegraphics{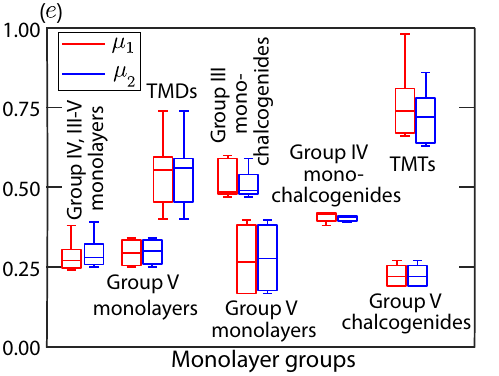}
\end{tocentry}

\begin{abstract}
We calculate transversal flexoelectric coefficients along the principal directions for fifty select atomic monolayers using ab initio Density Functional Theory (DFT). Specifically, considering representative materials from each of Groups IV, III-V, V monolayers, transition metal dichalcogenides (TMDs), Group III monochalcogenides, Group IV monochalcogenides, transition metal trichalcogenides (TMTs), and Group V chalcogenides, we perform symmetry-adapted DFT simulations  to calculate the flexoelectric coefficients at practically relevant bending curvatures. We find that the materials demonstrate linear behavior and have similar coefficients along both principal directions, with values for TMTs being up to a factor of five larger than graphene. In addition, we find electronic origins for the flexoelectric effect, which increases with monolayer thickness, elastic modulus along bending direction, and sum of  polarizability of constituent atoms. 
\end{abstract}
%%%END OF ABSTRACT%%%%
    
\rmfamily %Please do not remove this line.

%%%MAIN TEXT%%%%
%%%%%%%%%%%%%%%%%%%%

Flexoelectricity \cite{tagantsev1991electric, yudin2013fundamentals, zubko2013flexoelectric, nguyen2013nanoscale, ahmadpoor2015flexoelectricity, narvaez2016enhanced, krichen2016flexoelectricity, wang2019flexoelectricity} is an electromechanical property common to semiconductors/insulators that represents a two-way coupling between strain gradients and polarization. Unlike piezoelectricity, it is not restricted to materials that are non-centrosymmetric, i.e., lattice structures that do not possess inversion symmetry, and in contrast to electrostriction, it permits reversal of the strain by reversal of the electric field and allows for  sensing in addition to actuation. Though the flexoelectric  effect is generally negligible for bulk systems, it becomes  significant in nanostructures/nanomaterials due to the possibility of extremely large strain gradients, especially along the directions in which the system has dimensions at the nanoscale. 

The flexoelectric effect in two-dimensional materials has a number of applications --- analogous to those found for other such  electromechanical couplings \cite{hill2011graphene, wang1999electromechanical, xu2010self, wang2015observation, da2015strong} --- including sensors, actuators, and energy harvestors in nanoelectromechanical systems. Even in applications where the flexoelectric effect is not being exploited, e.g., flexible electronics \cite{pu2012highly, lee2013flexible, salvatore2013fabrication, yoon2013highly}, nanoelectromechanical devices \cite{zhang2015single, sakhaee2008potential, sazonova2004tunable, bunch2007electromechanical}, and nanocomposites \cite{novoselov2012two,qin2015lightweight},  the presence of strain gradients such as those encountered during bending  --- a common mode of deformation in two-dimensional materials, given their relatively low bending moduli values \cite{kumar2020bending} --- makes flexoelectricity an important design consideration \cite{2016-NANOSCALE-BBASRG, JAM2015-Abdollahi}. This is evidenced by recent work where flexoelectricity has been shown to produce incorrect measurements of piezoelectric coefficients at the nanoscale \cite{NatComm2019-Abdollahi}. 

Atomic monolayers, which are two-dimensional materials consisting of a single layer of material, have been the subject of intensive research over the past two decades \cite{mas20112d, butler2013progress, geng2018recent}, with  dozens of monolayers having now been synthesized \cite{balendhran2015elemental, zhou2018monolayer, vaughn2010single, zhang2018recent, dai2016group, wang2020bulk} and the potential for thousands more as predicted by ab initio calculations \cite{haastrup2018computational, zhou20192dmatpedia}. The widespread interest in these systems is a consequence of their novel and exciting properties \cite{balendhran2015elemental, zhou2018monolayer, zhang2018recent, kerszberg2015ab, fei2015giant, hsu2017topological,dai2016group, jiang2015recent}, which makes them ideal candidates for the aforementioned applications. However, the transversal flexoelectric coefficients for atomic monolayers --- the relevant component of the flexoelectric tensor in the context of bending deformations ---  are far from being established. 

Experimental data for the  transversal flexoelectric coefficients of atomic monolayers is highly sparse, likely due to the challenges associated with isolating the flexoelectric and piezoelectric contributions \cite{brennan2017out}. Recently, the coefficients for some TMDs (MX$_2$: M=Mo,W; X=S,Se) have been measured \cite{brennan2017out, brennan2020out}, however there is significant uncertainty in the results, due to large error bars and the use of substrates.  On the theoretical side, studies based on  ab initio Kohn-Sham DFT \cite{Hohenberg, Kohn1965} have been used to calculate the flexoelectric coefficients of graphene \cite{dumitricua2002curvature, nguyen2013nanoscale, kalinin2008electronic, shi2018flexoelectricity} and some TMDs (MX$_2$: M=Mo,W; X=S,Se,Te) \cite{shi2018flexoelectricity}. However, as shown in recent work \cite{codony2020flexo}, the accuracy of these results is limited by the use of an ill defined flexoelectric coefficient  \cite{kalinin2008electronic, shi2018flexoelectricity}, artificial partitioning of the electron density \cite{dumitricua2002curvature, kvashnin2015flexoelectricity}, and/or geometries with non-uniform strain gradients \cite{shi2018flexoelectricity}. Note that a theoretically more involved alternative is provided by density-functional perturbation theory (DFPT) \cite{gonze1997dynamical, baroni2001phonons} --- found to have significant success in the study of bulk like three-dimensional systems \cite{hong2011first, hong2013first, stengel2013flexoelectricity, stengel2014surface, dreyer2018current, royo2019first} --- which has very recently been extended to the study of two-dimensional systems \cite{springolo2020flexoelectricity}. Other theoretical efforts include the use of force fields \cite{javvaji2019high, zhuang2019intrinsic}, which differ by more than an order of magnitude from experimental/DFT results, suggesting that they are unsuitable in the current context.

In this work, using a recently developed formulation for the accurate computation of the transversal flexoelectric coefficient at finite deformations \cite{codony2020flexo}, we perform a comprehensive first principles DFT study for the  coefficients of fifty select atomic monolayers along their principal directions  at practically relevant bending curvatures \cite{lindahl2012determination, han2019ultrasoft, wang2019bending, qu2019bending}. We also provide fundamental insights into the flexoelectric effect for atomic monolayers and the variation in the coefficient values between them.

%%%%%%%%%%%%%%%%%%%%%%%%%%%%%%%

The transversal flexoelectric coefficient at finite bending deformations is  defined as: \cite{codony2020flexo}
\begin{equation} \label{Eq:FlexoCoefficient}
\mu = \frac{\partial p_r}{\partial \kappa} \,,
\end{equation}
where $p_r$ is the \emph{radial polarization} and $\kappa$ is the curvature. In the context of electronic structure calculations like DFT, the radial polarization can be expressed as: \cite{codony2020flexo} 
\begin{equation} \label{Eq:RadialPolarization}
p_r = \frac{1}{A} \int_{\Omega} (r-R_{\rm eff}) \rho(\mathbf{x}) \, \mathrm{d \mathbf{x}} \,,
\end{equation}
where $A$ is the cross-sectional area of the deformed sheet within the domain $\Omega$, $r$ is the radial coordinate of the  spatial point $\mathbf{x}$, $R_{\rm eff}$ is the \emph{radial centroid} of the ions, and $\rho(\mathbf{x})$ is the electron density.  Note that the \emph{radial dipole moment} has been normalized using the area rather than the volume,  as is common practice, given the significant disagreement associated with the thickness of atomic monolayers \cite{huang2006thickness}. 

\begin{figure}[htbp!]
\includegraphics[width=0.99\textwidth]{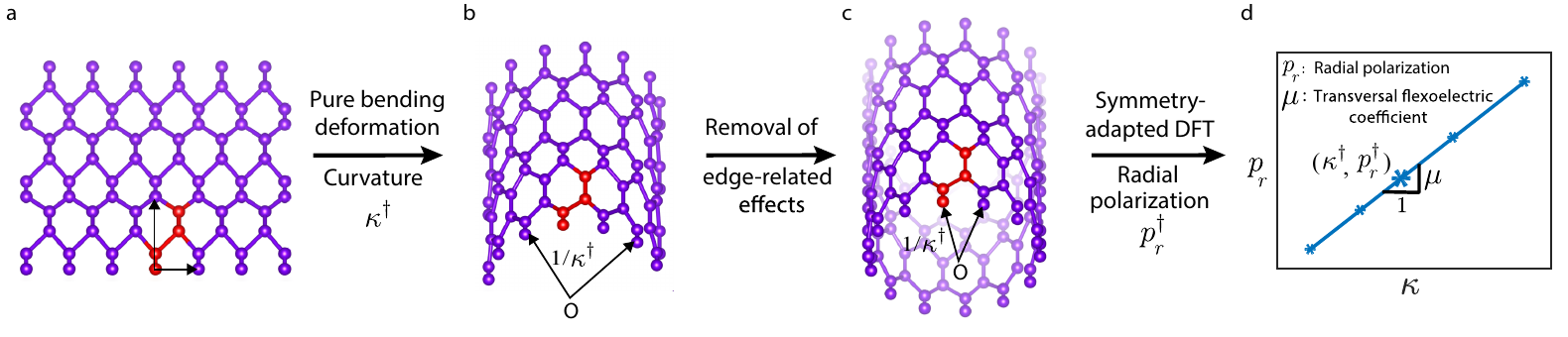}
\centering
\caption{Schematic illustrating the calculation of transversal flexoelectric coefficient for atomic monolayers using symmetry-adapted DFT simulations \cite{codony2020flexo, sharmaCyclix,  ghosh2019symmetry}. The atoms in the unit cell are colored red. }
\label{fig:illustration}
\end{figure}

We calculate the flexoelectric coefficient by using a numerical approximation for the derivative in Eq.~\ref{Eq:FlexoCoefficient}, the alternative being the more involved DFPT-based approaches, for which a symmetry-adapted variant at finite deformations is yet to be developed.  Specifically, we compute the radial polarization at multiple curvatures in the vicinity of the curvature at which the flexoelectric coefficient is desired and  determine the slope from a curve fit to the data. In particular, as illustrated in Fig.~\ref{fig:illustration}, edge-related effects are removed by mapping the bent structure periodically in the angular direction, and the cyclic symmetry of the resultant structure is then exploited  to perform highly efficient Kohn-Sham calculations \cite{sharmaCyclix,ghosh2019symmetry, banerjee2016cyclic} for the structural and electronic ground state using the Cyclix DFT code \cite{sharmaCyclix} --- large-scale parallel implementation of cyclic+helical symmetry in state-of-the-art real-space DFT code SPARC  \cite{xu2020sparc, ghosh2017sparc2}. 

We use the aforedescribed framework to calculate transversal flexoelectric coefficients for fifty select atomic monolayers along their principal directions. Specifically, we consider bending curvatures in the range of $0.14<\kappa<0.24$ nm$^{-1}$ --- resulting system sizes are intractable to traditional DFT implementations --- commensurate with those found in experiments \cite{lindahl2012determination, han2019ultrasoft, wang2019bending, qu2019bending}.  We select representative honeycomb lattice materials  from each of Groups IV, III-V, V monolayers, TMDs, and Group III monochalcogenides, as well as rectangular lattice materials from each of Group V monolayers, Group IV monochalcogenides, TMTs, and Group V chalcogenides. The choice of these groups is motivated by the significant success in the synthesis of affiliated monolayers, which are found to demonstrate interesting and novel properties \cite{dai2016group, Coleman568, Novoselov10451, zhou2018monolayer, vaughn2010single, zhang2018recent, balendhran2015elemental, siskins2019highly}.  

In all simulations, we employ the Perdew-Burke-Ernzerhof (PBE) \cite{perdew1986accurate}  variant of the generalized gradient approximation (GGA) for the exchange-correlation functional and optimized norm-conserving Vanderbilt (ONCV) pseudopotentials \cite{hamann2013optimized} from the SG15 collection \cite{SCHLIPF201536}. All numerical parameters, including  grid spacing, k-point sampling for Brillouin zone integration, vacuum in radial direction, and structural (cell and atom) relaxation tolerances are chosen such that the computed flexoelectric coefficients are accurate to within $0.005\,e$, as verified through convergence studies (Supplementary Material). This translates to the requirement of the ground state energy being converged to within $10^{-5}$ Ha/atom. Note that the coefficients predicted here are expected to be reasonably robust against the two main approximations within DFT, i.e., pseudopotential and exchange-correlation functional, as dicussed next. 

The  equilibrium geometries for the flat monolayers computed using ABINIT \cite{ABINIT} (Supplementary Material)  are  in good agreement with previous theoretical predictions \cite{zhou20192dmatpedia, haastrup2018computational} and experimental measurements \cite{dai2016group, Coleman568, Novoselov10451, zhou2018monolayer, vaughn2010single, zhang2018recent, balendhran2015elemental}. Furthermore, the normalized difference in electron density between the PBE GGA and HSE hybrid functional \cite{heyd2003hybrid} for the undeformed configurations is  $\mathcal{O} (1-2)$,  comparable to the agreement between hybrid functionals and the gold standard Coupled Cluster Singles and Doubles (CCSD) method \cite{medvedev2017density}. Similar differences are observed when spin orbit coupling is included.  Since the flexoelectric coefficient is dependent on electron density differences from the flat configuration, which corresponds to small (linear) perturbations in the current context, significant error cancellations are expected. This is evidenced by recent work where both local and semilocal functionals predict nearly identical coefficients for group IV monolayers  \cite{codony2020flexo}.

\begin{table}[htbp!]
    \centering
    \caption{Transversal flexoelectric coefficient along principal directions for the fifty select atomic monolayers from first principles DFT calculations.}
    \label{Table:FC}
    %\begin{indented}
    \resizebox{1.0\textwidth}{!}{
    \begin{tabular}{|c|c|c|c|c|c|c|c|}
   \hline
   \multirow{3}{*}{Group} &\multirow{3}{*}{Material}&\multicolumn{2}{c|}{Flexoelectric}&\multirow{3}{*}{Group}&\multirow{3}{*}{Material}&\multicolumn{2}{c|}{Flexoelectric}\\
    &&\multicolumn{2}{c|}{coefficient ($e$)}&&&\multicolumn{2}{c|}{coefficient ($e$)}\\
 
    & &\multicolumn{1}{c}{ $\mu_1$} &\multicolumn{1}{c|}{$\mu_2$} &&  & \multicolumn{1}{c}{$\mu_1$} & \multicolumn{1}{c|}{$\mu_2$}\\
    \hline
    \multirow{2}{*}{ Groups IV, III-V}& Si& 0.19 & 0.19 &&GaS& 0.48 & 0.48 \\
    \cline{2-4}
    \cline{6-8}
    \multirow{2}{*}{monolayers}& BN& 0.20 & 0.20& \multirow{2}{*}{Group III} & GaSe & 0.48 & 0.50  \\
    \cline{2-4}
    \cline{6-8}
    \multirow{2}{*}{(h1)}& C & 0.22 & 0.22 &\multirow{2}{*}{monochalcogenides} & InS & 0.47 & 0.47 \\
    \cline{2-4}
    \cline{6-8}
    & Sn& 0.26 & 0.25  &\multirow{2}{*}{(h3)}& InSe&0.49 &0.48\\
    \cline{2-4}
    \cline{6-8}
    & Ge & 0.27 & 0.27 &  & InTe & 0.59 & 0.54\\
    \cline{1-4}
    \cline{6-8}
    \multirow{2}{*}{Group V}&P&  0.25& 0.25 & &GaTe & 0.60& 0.59  \\
    \cline{2-4}
    \cline{5-8}
     \multirow{2}{*}{monolayers}&As&  0.26& 0.27&  \multirow{2}{*}{Group V}&P& 0.31 &0.33\\
    \cline{2-4}
    \cline{6-8}
     \multirow{2}{*}{(h1)}&Bi &  0.33& 0.33& \multirow{2}{*}{monolayers} &As& 0.31& 0.31\\
    \cline{2-4}
    \cline{6-8}
    &Sb &  0.34& 0.34 &\multirow{2}{*}{(t1)}&Bi& 0.51 & 0.51\\
    \cline{1-4}
    \cline{6-8}
     & ZrS$_2$ &  0.45 & 0.45 & &Sb& 0.54 & 0.54 \\
    \cline{2-4}
    \cline{5-8}
    & TiS$_2$ &  0.45 & 0.45 & \multirow{2}{*}{Group IV}&GeSe & 0.38& 0.39 \\
    \cline{2-4}
    \cline{6-8}
    &ZrSe$_2$ & 0.46 & 0.47 &\multirow{2}{*}{monochalcogenides}&GeS & 0.42& 0.41 \\
    \cline{2-4}
    \cline{6-8}
    &  TiSe$_2$& 0.41&0.40  & 
    \multirow{2}{*}{(t1)} &SnSe  &0.41 &0.41 \\
    \cline{2-4}
    \cline{6-8}
      &NbS$_2$ & 0.51 &  0.52 & &SnS & 0.42&0.40 \\
    \cline{2-4}
    \cline{5-8}
     & NbSe$_2$& 0.51 & 0.54 &  &ZrS$_3$ &0.66&0.64\\
    \cline{2-4}
    \cline{6-8}
    \multirow{1}{*}{Transition metal}& HfS$_2$ &0.57 & 0.56 & &TiS$_3$&0.67&0.63\\
    \cline{2-4}
    \cline{6-8}
    \multirow{1}{*}{dichalcogenides}& ZrTe$_2$&  0.55 & 0.54& \multirow{1}{*}{Transition metal}& ZrSe$_3$&0.68 &0.66\\
    \cline{2-4}
    \cline{6-8}
   \multirow{1}{*}{(h2)}& TiTe$_2$ & 0.58& 0.56&\multirow{1}{*}{trichalcogenides} &HfS$_3$  &0.80& 0.78\\
    \cline{2-4}
    \cline{6-8}
     & MoSe$_2$& 0.57& 0.57&  \multirow{1}{*}{(t2)}  &HfSe$_3$ &0.81 & 0.78\\
    \cline{2-4}
    \cline{6-8}
     & MoS$_2$ & 0.57& 0.58 & & ZrTe$_3$& 0.98 & 0.86\\
    \cline{2-4}
    \cline{5-8}
     & WS$_2$ & 0.59 &  0.59&  &  P$_2$S$_3$ &0.24& 0.25\\
    \cline{2-4}
    \cline{6-8}
     &WSe$_2$ &0.59&  0.58 &  \multirow{2}{*}{Group V}& P$_2$Se$_3$ & 0.25& 0.26\\

     \cline{2-4}
    \cline{6-8}
      & NbTe$_2$ &0.64&  0.67 & \multirow{2}{*}{chalcogenides} & As$_2$S$_3$ & 0.27& 0.28\\
      \cline{2-4}
    \cline{6-8}
      & MoTe$_2$ &0.71&  0.72 & \multirow{2}{*}{(t3)} & As$_2$Se$_3$ & 0.28& 0.30\\
      \cline{2-4}
    \cline{6-8}
      & WTe$_2$ &0.73&  0.74 &  & As$_2$Te$_3$ & 0.38& 0.39\\
      \hline
    \end{tabular}
    }
  % \end{indented}
   
\end{table}

\begin{figure}[htbp!]
	    \centering
		\includegraphics[width=0.44\textwidth]{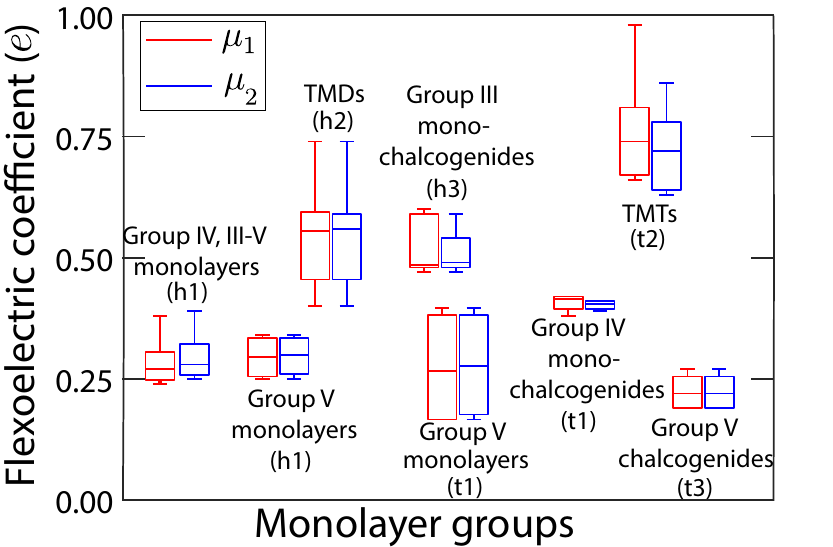}
		\caption{Transversal flexoelectric coefficients for the select atomic monolayer groups.}
		\label{fig:boxplot}
\end{figure}
%%%%%%%%%%%%%%%

\begin{figure}[htbp!]
	    \centering
		\includegraphics[width=0.9\textwidth]{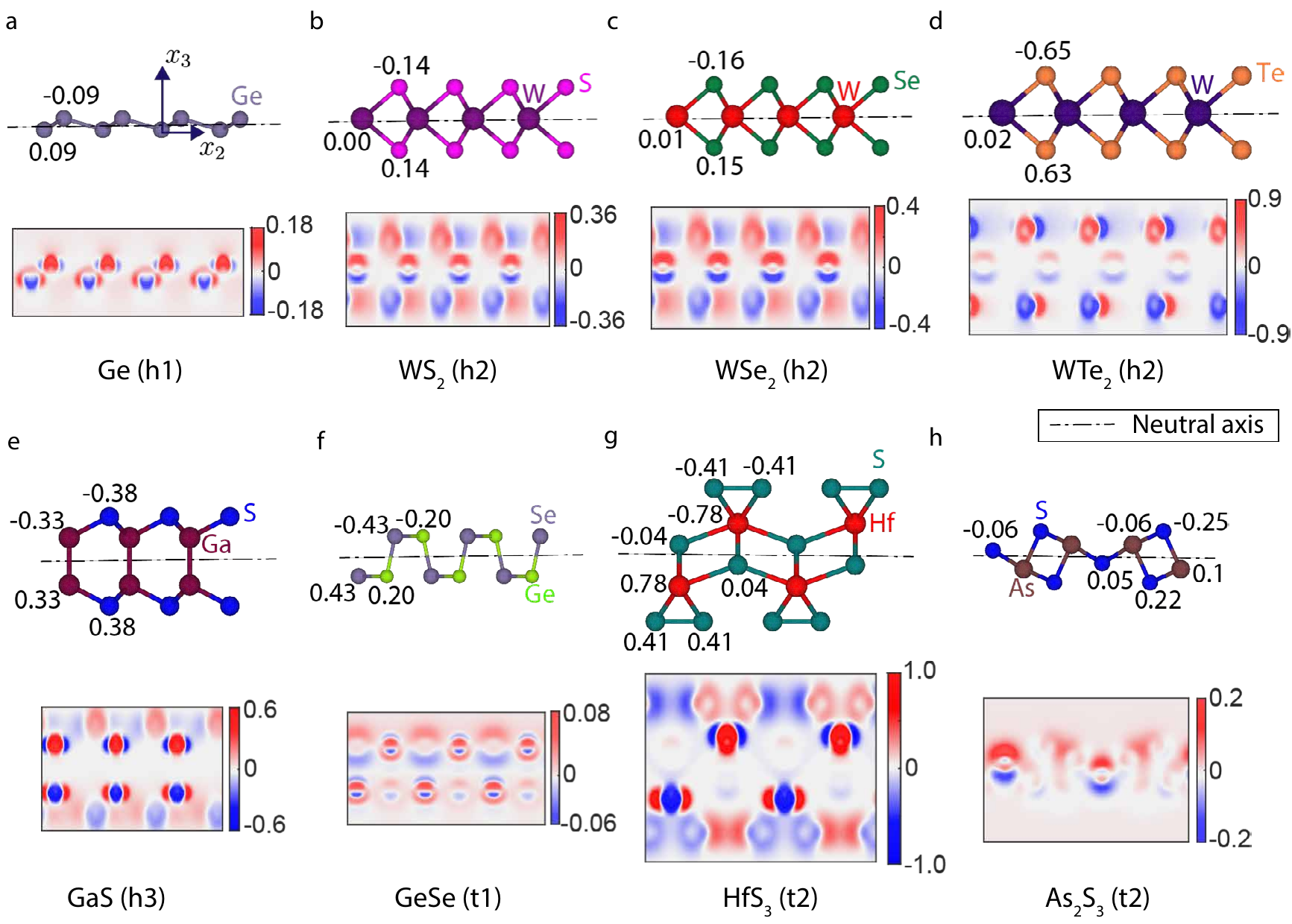}
		\caption{Contours of electron density difference (integrated along the $x_1$ direction) between the flat and bent ($\kappa$ $\sim$ 0.2 nm$^{-1}$) atomic monolayers. The contours are plotted on the $x_2 x_3$-plane in the undeformed configuration. The charge transfer due to bending, which is shown near the corresponding atoms in the lattice structure, is obtained from Bader analysis \cite{bader1981quantum}.}
		\label{fig:struc:cont}
\end{figure}

In Table~\ref{Table:FC}, we present the computed transversal flexoelectric coefficients for the chosen atomic monolayers along their principal directions, which is summarized visually in Fig.~\ref{fig:boxplot}. The variables $\mu_1$ and $\mu_2$ are used to represent the flexoelectric coefficient values along the $x_1$ and $x_2$ directions, respectively, whose orientation relative to the different lattice structures can be seen in Fig.~\ref{fig:struc:cont}. For honeycomb lattices, these correspond to the zigzag and armchair directions, respectively. A single value is listed in all cases since the flexoelectric coefficients are essentially constant for the bending curvatures considered here (Supplementary Material), signaling linear response for the monolayers in this regime. Note that depending on the application of interest, the flexoelectric coefficient that relates polarization to bending moment might be more informative. Therefore, the values for the so defined flexoelectric coefficient have been provided in the Supplementary Material, for which we use the bending moduli   from previous work for forty-four of the materials \cite{kumar2020bending}, with the remaining calculated here (Supplementary Material).

The flexoelectric coefficients span a wide range of values from $0.19 - 0.98 \, e$, with silicene and  ZrTe$_3$ being at the bottom and top ends of the spectrum, respectively, and graphene towards the lower end with $0.22\,e$. In terms of the classification, Groups IV, III-V monolayers and TMTs have the smallest and largest coefficients, respectively. Interestingly, we find that the flexoelectric coefficients are similar along both principal directions, irrespective of the lattice structure. Though this is to be expected for honeycomb lattices, which usually demonstrate isotropic behavior/properties \cite{kumar2020bending, ding2019most, rasmussen2015computational, li2015piezoelectricity}, it is most unusual for rectangular lattices, where the behavior/properties tend to be highly anistropic \cite{kumar2020bending, li2019emerging, xia2014rediscovering, wang2015highly, luo2015anisotropic}. This is not a consequence of relaxation-related effects --- cause for the bending moduli of some rectangular lattices to be isotropic \cite{kumar2020bending} --- which are minor in the current context (Supplementary Material). Note that when the flexoelectric coefficient relating the polarization to the bending moment is considered (Supplementary Material), the values span more than two orders of magnitude, with the trends essentially reversed, i.e., Groups IV, III-V monolayers and TMTs now have the largest and smallest coefficients, respectively. In particular, stanene has the largest value, given its extremely small bending moduli \cite{kumar2020bending}, and ZrTe$_3$ has the smallest value, given its extremely large bending moduli \cite{kumar2020bending}. Also, the coefficients for the rectangular lattices differ significantly in the principal directions, given their significant anisotropy in bending moduli \cite{kumar2020bending}.

In comparisons with experiments, while there is relatively good agreement for MoS$_2$ (difference of $\sim 0.09 \, e$ \cite{brennan2017out}), there is some disagreement for MoSe$_2$, WS$_2$, and WSe$_2$ (differences of up to $\sim 0.48 \, e$ \cite{brennan2020out}). These differences can be attributed to the use of a substrate,  and the substantial error  bars ($\sim 0.18 \, e$) associated with the measurements. In comparisons to  DFT-based  results, the values for graphene predicted previously \cite{kalinin2008electronic, dumitricua2002curvature, kvashnin2015flexoelectricity} are more than a factor of two smaller than those here. This can be attributed to the use of an ill defined  flexoelectric coefficient \cite{kalinin2008electronic} and an artificial partitioning of the electron density \cite{dumitricua2002curvature, kvashnin2015flexoelectricity}, as discussed in previous work \cite{codony2020flexo}. The values  reported previously  for TMDs \cite{shi2018flexoelectricity} (MX$_2$: M=Mo,W; X=S,Se,Te) are up two orders of magnitude smaller than those here, a consequence of using an ill defined flexoelectric coefficient and a wrinkled sheet geometry that has non-uniform strain gradients. Similarly small values for a few monolayers have been predicted  very recently  \cite{springolo2020flexoelectricity}, which can be attributed to the difference in the definition of the flexoelectric coefficient. 

%%%%%%%%%%%%%%%
To get further insight into the results, considering representative materials from each group, we plot in Fig.~\ref{fig:struc:cont} contours of electron density difference between the flat and bent monolayers for bending along the $x_2$ direction. We also present the charge transfer due to bending as determined via Bader analysis \cite{bader1981quantum}. Similar results for bending along the $x_1$ direction can be found in the Supplementary Material. Three key observations can be made from these figures.  First, charge transfer occurs from the compressive side to tensile side of the neutral axis, indicating that the origin of the flexoelectric effect for monolayers is electronic rather than ionic. The charge transfer is similar for both bending directions, resulting in similar flexoelectric coefficients. Second, the electron density perturbations are localized near the nuclei, resulting in atomic dipoles that accumulate to generate the total radial dipole moment. The strength of these dipoles is dependent on the atom's polarizability,  as evidenced from the significantly larger charge transfer for Te compared to S and Se in the tungsten dichalcogenides. Indeed, polarizability of S and Se are similar, both of which are significantly smaller than Te \cite{tandon2019new}.  Third, the atomic polarization (i.e., atomic dipole per unit area) increases with distance from neutral axis, as evidenced  for the S atom in the various monolayers. This can be attributed to the increase in stress with neutral axis distance.

\begin{figure}[htbp!]
	    \centering
		\includegraphics[width=0.99\textwidth]{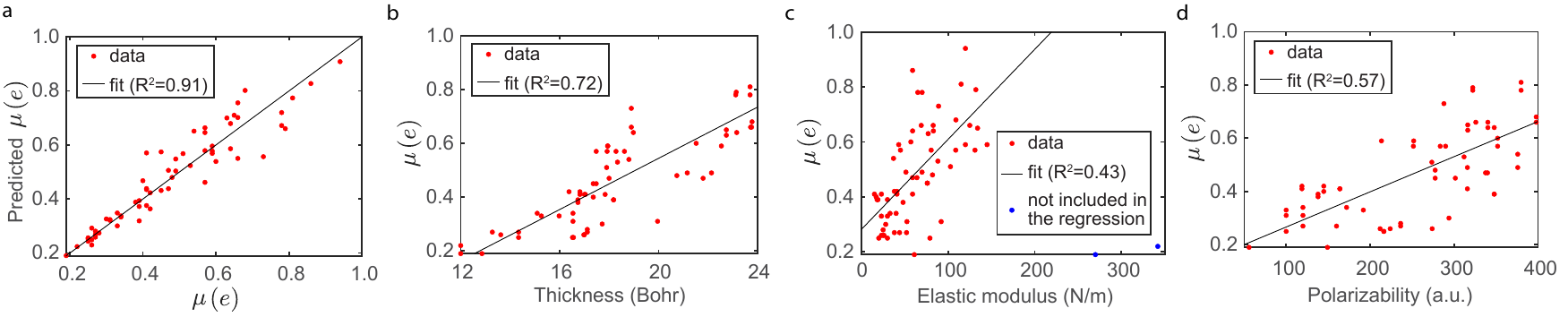}
		\caption{(a) Set of calculated transversal flexoelectric coefficients and its third order polynomial regression with the features being thickness, elastic modulus along bending direction, and sum of polarizability of constituent atoms. (b), (c), and (d) Set of computed flexoelectric coefficients and its linear regression with the feature being thickness,  elastic modulus along bending direction, and sum of polarizability of constituent atoms, respectively. In all plots, $R^2$ denotes the coefficient of determination for the regression.}
		\label{fig:struc:regression}
\end{figure}

The above observations suggest the flexoelectric coefficients for the monolayers is primarily determined by their thickness, elastic modulus along bending direction, and sum of polarizabilities of constituent atoms. The dependence on atom polarizabilities is in agreement with  literature \cite{tagantsev1985theory, Ma2019}, where it has been proposed that the flexoelectric coefficient is proportional to the dielectric permittivity, which can be related to the atom polarizabilities through the Clausius–Mossotti relation \cite{griffiths2005introduction}. Using the three features listed above, we perform a third order polynomial regression for the  flexoelectric coefficients, the results of which are presented in Fig.~\ref{fig:struc:regression}. Note that the thickness  has been defined to be the distance between the two atoms furthest from the neutral axis plus an additional 12 Bohr (results  insensitive to this choice). The fit is very good, suggesting that the flexoelectric coefficients for atomic monolayers are primarily decided by  the three aforementioned features. We also perform a linear regression between the computed coefficients and each of the features independently, the results of which are presented in Fig.~\ref{fig:struc:regression}. The fits are good, suggesting that the  flexoelectric coefficient generally increases with monolayer thickness, elastic modulus along bending direction, and sum of  polarizability of constituent atoms. 

In summary, we have calculated transversal flexoelectric coefficients along the principal directions for fifty select atomic monolayers using ab initio DFT. Specifically, considering representative materials from each of the prominent monolayer groups, we have determined the coefficients at practically relevant bending curvatures using symmetry-adapted DFT calculations. We have found that the monolayers demonstrate linear behavior and have similar flexoelectric coefficients along both principal directions, with values for TMTs being up to a factor of five larger than graphene. In addition, we have found electronic origins for the flexoelectric effect, which increases with monolayer thickness, elastic modulus along bending direction, and sum of  polarizability of constituent atoms. Overall, this work provides an important reference for the transversal flexoelectric coefficients for a number of important  atomic monolayers, and provides fundamental insights into the underlying mechanisms. The flexoelectric coefficients predicted here could prove useful in the design of nanoelectromechanical devices, with the regression model serving as a powerful tool for preliminary searches through large databases of two-dimensional materials.

\section*{Acknowledgments}
S.K. and P.S. gratefully acknowledge the support of the U.S. National Science Foundation (CAREER–1553212 and MRI-1828187). D.C. and I.A. acknowledge the support of the Generalitat de Catalunya (ICREA Academia award for excellence in research to I.A., and Grant No. 2017-SGR-1278), and the European Research Council (StG-679451 to I.A.). CIMNE is recipient of a Severo Ochoa Award of Excellence from the MINECO.

\section*{Supporting Information Available}
Equilibrium geometry for the atomic monolayers, Unrelaxed transversal flexoelectric coefficients,  Radial polarization vs. curvature data, Convergence study for the computed flexoelectric coefficients, Flexoelectric coefficient relating polarization to bending moment,  Bader analysis results for bending along the $x_1$-direction, Bending moduli for select monolayers

\section*{Conflicts of interest}
There are no conflicts to declare.

%%%%%%%%%%%%%%%%%%%%%%%%%%%%%%%%%%%%%%	

%\bibliography{Manuscript} 

\providecommand{\latin}[1]{#1}
\makeatletter
\providecommand{\doi}
  {\begingroup\let\do\@makeother\dospecials
  \catcode`\{=1 \catcode`\}=2 \doi@aux}
\providecommand{\doi@aux}[1]{\endgroup\texttt{#1}}
\makeatother
\providecommand*\mcitethebibliography{\thebibliography}
\csname @ifundefined\endcsname{endmcitethebibliography}
  {\let\endmcitethebibliography\endthebibliography}{}

%%%%%%%%%%%%%%%%%%%%%%%%%%%%%%%%%%%%%%

\end{document}